\documentclass{pasj00}
\begin{document}

\SetRunningHead{H. Takami et al.}{Running Head}
\Received{2007/05/05}
\Accepted{2009/03/19}

\title{Direct Observation of the Extended Molecular Atmosphere of $o$~Cet by Differential Spectral Imaging with an Adaptive Optics System}
\author{%
Hideki \textsc{Takami}\altaffilmark{1}
Miwa \textsc{Goto}\altaffilmark{2}
Wolfgang \textsc{Gaessler}\altaffilmark{2}
Yutaka \textsc{Hayano}\altaffilmark{1}
Masanori \textsc{Iye}\altaffilmark{3}
Yukiko \textsc{Kamata}\altaffilmark{3}
Tomio \textsc{Kanzawa}\altaffilmark{1}
Naoto \textsc{Kobayashi}\altaffilmark{4}
Yosuke \textsc{Minowa}\altaffilmark{1}
Shin \textsc{Oya}\altaffilmark{1}
Tae-Soo \textsc{Pyo}\altaffilmark{1}
David \textsc{Saint-Jacques}\altaffilmark{5}
Naruhisa \textsc{Takato}\altaffilmark{1}
Hiroshi \textsc{Terada}\altaffilmark{1}
Alan T. \textsc{Tokunaga}\altaffilmark{6}
and Takashi \textsc{Tsuji}\altaffilmark{4}
}

\altaffiltext{1}{Subaru Telescope, 650 North A`ohoku Place, Hilo,
HI 96720.}
\email{takami@naoj.org}

\altaffiltext{2}{Max-Planck-Institut f\"ur Astronomie, K\"onigstuhl 17,
Heidelberg D-69117, Germany.} 

\altaffiltext{3}{National Astronomical Observatory of Japan, Osawa,
Mitaka, Tokyo 181-8588, Japan.} 

\altaffiltext{4}{Institute of Astronomy, The University of Tokyo, Osawa,
Mitaka, Tokyo 181-0015, Japan.} 

\altaffiltext{5}{Groupe d'astrophysique, Universit$\acute{e}$ de Montr$\acute{e}$al, 2900 Boul.
$\acute{E}$douard-Montpetit, Montr$\acute{e}$al (QC) H3C 3J7, Canada}

\altaffiltext{6}{Institute for Astronomy, University of Hawaii, 2680
Woodlawn Dr., Honolulu, HI 96822.}


%

\KeyWords{ infrared: general, instrumentation: adaptive optics, stars: AGB and post-AGB, techniques: spectroscopic} 

\maketitle

\begin{abstract}
We present new measurements of the diameter of $o$~Cet (Mira) as a function of wavelength in the 2.2~\micron~atmospheric window using the adaptive optics system and the infrared camera and spectrograph mounted on the Subaru Telescope. We found that the angular size of the star at the wavelengths of CO and H$_2$O absorption lines were up to twice as large as the continuum photosphere. This size difference is attributable to the optically thick CO and H$_2$O molecular layers surrounding the photosphere. This measurement is the first direct differential spectroscopic imaging of stellar extension that resolves individual molecular lines with high spectral-resolution observations. 
This observation technique is extremely sensitive to differences in spatial profiles at different wavelengths; we show that a difference in diameter much smaller than the point spread function can be measured. 

\end{abstract}

\section{Introduction}

Mira-type stars are important in studying galactic evolution because the gases and dust ejected from them constitute a major contribution to galactic heavy element enrichment. However, detailed study of the physical structure of the atmosphere of Mira-type stars was not feasible until recently, because the angular size of these stars is smaller than the resolution achieved using seeing-limited optical and infrared telescopes. Although their deep absorption features are clear evidence of the presence of molecular gases such as H$_2$O and CO, their spatial distribution in the outer atmosphere is not well understood. 

Recent observations using satellites and ground-based interferometers have greatly improved our knowledge of the spatial distribution of molecular gases. High spatial-resolution observations using infrared interferometers have been used to study stellar diameters \citep{rid92, dyc96, dyc98}. \citet{men02} compared the stellar size of cool M giants at $K'$ and $L'$ and found that stellar diameters are larger at longer wavelengths. They attributed this to emissions from an extended, partially transparent shell, although the source of the opacity was not identified. \citet{ire04} showed that measured stellar diameters are larger at molecular absorption bands in the near infrared region. \citet{per04} reported that the diameter is larger at the 2.39~\micron~band due to water vapor and CO molecules, and that molecular layers extend out to 2.2 $R_{\rm \ast}$. 

Based on spectroscopic observations using the Infrared Space Observatory (ISO), \citet{yam99} modeled the outer atmosphere of Mira, a representative late M-type star, which has a "hot" H$_2$O shell at 2.0 $R_{\rm \ast}$ with a column density of 3 $\times$ $10^{21}$ cm$^{-2}$ and a temperature of 2000~K, and a "cool" H$_2$O shell at 2.3 $R_{\rm \ast}$ characterized by a column density of 3 $\times$ $10^{20}$ cm$^{-2}$ and a temperature of 1400~K. Here $R_{\rm \ast}$ denotes the radius of the stellar photosphere. 
\citet{mat02} shows that the very extended water vapor layer is a common feature in the Mira variables. 

Working from near-infrared (NIR) to mid-infrared (MIR) interferometric observation data, \citet{ohn04} presented a two-layer model consisting of hot and cool
H$_2$O layers for three Mira variables and concluded that they have a hot 1800-2000~K layer between 1.5 and 1.7 $R_{\rm \ast}$ with a column density of 1 to 5 $\times$ $10^{21}$ cm$^{-2}$, and a cool 1200-1400~K layer between 2.2 and 2.5 $R_{\rm \ast}$ with a column density of 1 to 7 $\times$ $10^{21}$ cm$^{-2}$. \citet{wei04} explained the enhancement of stellar size from NIR to MIR using a 2000~K water vapor shell model. 
 
Dynamic modeling of the pulsation of Mira variables shows that the atmosphere extends well outside the photosphere where temperature is low enough to form molecules \citep{hoe98}. 
Attempts to identify the molecular species contributing to the extension of the photosphere using multiwavelength interferometry in the $J$, $H$, and $K$ bands were conducted with the Infrared Optical Telescope Array(IOTA) \citep{wei03}. Observations with narrow band filters in the $K$ band were carried out with the Palomar Testbed Interferometer \citep{tho02}. However, the spectral resolution was not sufficient to resolve molecular lines. Recently, the Keck interferometer (spectral resolution R = 230) was able to observe the CO and water vapor bands at 2.0-2.4~\micron \citep{eis07}. While those lines were not clearly resolved at that spectral resolution, their observations start to show the molecular features contributing to the extension of the photosphere. The exposure time for interferometry has to be very short, and the telescope aperture sizes are typically small. Therefore, it is difficult to increase the spectral resolution of an interferometer while maintaining a sufficiently high signal-to-noise ratio.
 
In this paper, we use an instrument featuring adaptive optics (AO) correction that is suited to studying extended structures around a compact source with a spectral resolution sufficiently high that it can resolve individual molecular lines. The diffraction-limited point spread function (PSF) of 8-m class telescopes is 32-56~mas at 1.25-2.2~\micron, while the largest stellar disk angular size spanned by the nearby red super-giant $a$~Ori is 44.2~mas at 2.2~\micron~ \citep{dyc92}. Therefore, the spatial resolution is insufficient to measure its stellar diameter by direct imaging under variable seeing conditions. Nevertheless, we can derive the wavelength dependence of the stellar diameter by conducting spectroscopic observations with AO. Diffraction-limited spectroscopy provides one-dimensional imaging of the stellar size at hundreds of spectral resolution elements recorded simultaneously. By comparing the spatial profiles at some line features to those at the immediately adjacent continuum wavelength, one can achieve very sensitive measurements of the stellar diameter because the PSF variation with time is essentially common and eliminated. We used this {\it differential spectral imaging} technique to investigate the stellar size in and out of the molecular lines to observe the extended molecular atmosphere of the late-type star $o$~Cet directly.

The observation procedure used in differential spectral imaging is essentially the same as that of classical dispersion spectroscopy. Some issues, however, must be carefully considered. First, the correction by AO must be sufficiently good to deliver a PSF with a clear diffraction-limited core. Second, observations should be carried out on low-humidity nights because the water vapor in the stellar atmosphere is one of our target molecules and the PSF depends on the refraction indices of the telluric atmosphere, which vary considerably at these strong absorption lines. In our experience, as long as the humidity is below 10\%, which is not uncommon at Mauna Kea, the effect of the wavelength dependency of the refraction index is negligible.

\section{Observation}
$o$~Cet (Mira, M5e-M9e, 332-day variability period; \cite{kho98}) is one of the brightest late-type M giants. The outer structure of the star has been studied using space-based spectroscopy \citep{yam99} and interferometry at optical and IR wavelengths \citep{rid92, han95, woo04}. We conducted spectroscopic imaging of the object using the 36-element curvature-based adaptive optics system of the Subaru Telescope to achieve diffraction-limited resolution at 2~micron~\citep{tak04, gae02}.

The observations were made on UT 2001 December 22 during the commissioning run of the AO system using the Infrared Camera and Spectrograph (IRCS) \citep{tok98, kob00}. The night was ideal for the planned observation as the humidity was low and the seeing conditions were favorable. We chose the $K$ band covering the absorption bands from two abundant molecules, CO at 2.3~\micron~and H$_2$O at 2.45~\micron.

A medium-resolution grism in the IRCS with 23~mas/pixel was used with a 0\farcs10 slit to provide spectra at a resolving power of 1000. The position angle of the slit was set along the east--west direction. The object itself served as a wavefront sensing reference for the AO system. The spectrograms were recorded by chopping the AO tip-tilt mirror by 3\arcsec~along the slit. To avoid detector saturation from the very bright central object, we used a neutral density filter, and set the integration time to 0.26~s for a single exposure. One hundred exposures were co-added before readout to storage. The total time from the start of the first exposure to the end of the last exposure was 26~s including read-out after each integration, which is sufficiently long to average out the speckle noise. A comparison star of similar spectral type (SW~Cet, M7~III; \cite{kho98}) was also observed. SW~Cet is 60 times fainter than $o$~Cet at $K$ and is used as a reference point source, since it is located eight times farther away from us than $o$~Cet. An early type spectroscopic standard star was observed through a similar air mass as the program star to remove the telluric absorption lines. The spectroscopic flat fields were obtained at the end of the night with a halogen lamp.

\section{Data Analysis and Discussion}
We subtracted pairs of frames that had the spectral images positioned 3" apart from each other on the IR array detector to remove the sky emission and dark-count patterns. The spatial profile at each spectral element along the slit was fitted with a three-component Gaussian function to represent the diffraction core, the shoulder shape of the first Airy ring, and the seeing halo. This is the typical PSF profile with partially corrected AO observation; therefore, we have introduced those components to our fitting calculation. There are no fixed parameters in the fitting. The free parameters are the intensity, spectral width, and the wavelength for all three components.
The full-width at half-maximum (FWHM) of the spatial extent was measured against the synthetic profile at each wavelength. The spectra are slightly tilted along the pixel row by as much as 1\%. However, rectification using a sub-pixel shift was not performed to avoid resampling effects on the spatial profile. The effect of non-rectification is negligible. If it has any effect, the measured FWHM should have a periodic pattern at about a 100-pixel interval. However, we did not observe such a pattern in the derived FWHM. 

The spectra were extracted using the IRAF\footnote{IRAF is distributed by the National Optical Astronomy Observatories, which are operated by the Association of Universities for Research in Astronomy, under a cooperative agreement with the National Science Foundation.} aperture extraction package. The wavelength was calibrated by maximizing the cross-correlation of the spectra with the modeled telluric atmospheric absorption curve calculated using the program ATRAN \citep{lor92}.

\begin{figure}[p]
 \begin{center}
 \FigureFile(80mm, 80mm){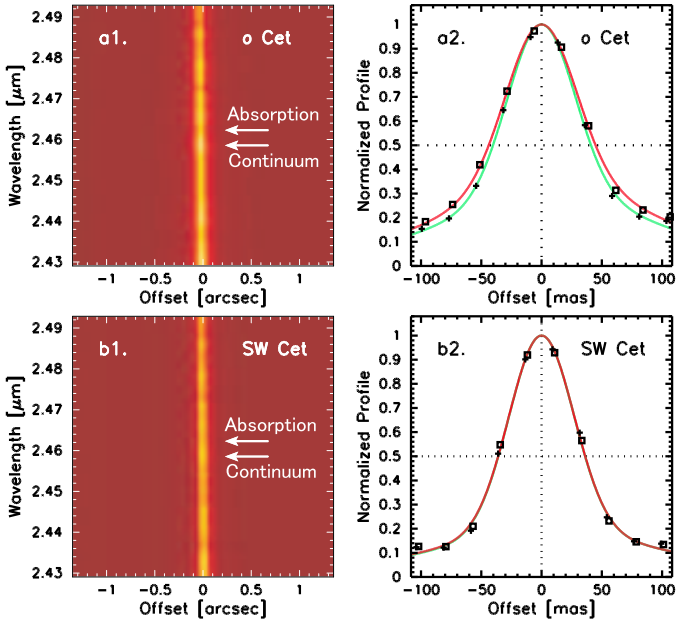}
 \end{center}
 \caption{(a1) Blow-up of the spectrogram of $o$~Cet near 2.4~\micron. The arrows point to a stellar absorption line and an adjacent continuum region. (a2) Comparison of the measured spatial profiles, rectangles and crosses, extracted at the absorption line and at the continuum region marked in (a1), respectively. The spectra appears wider at the bright continuum wavelength. However, if we normalize to the peak intensity, the stellar size extends more at the molecular line wavelength. The solid red and green curves are three-component Gaussian functions fitted to the respective data points. The spatial profile at the absorption line is wider than that at the continuum region. (b1), (b2) Same as (a1) and (a2), but for the PSF reference star SW~Cet, which is eight times farther than $o$~Cet and can be regarded as an unresolved point source. \label{f1}}
\end{figure}

\begin{figure}[p]
\begin{center}
 \FigureFile(80mm,70mm){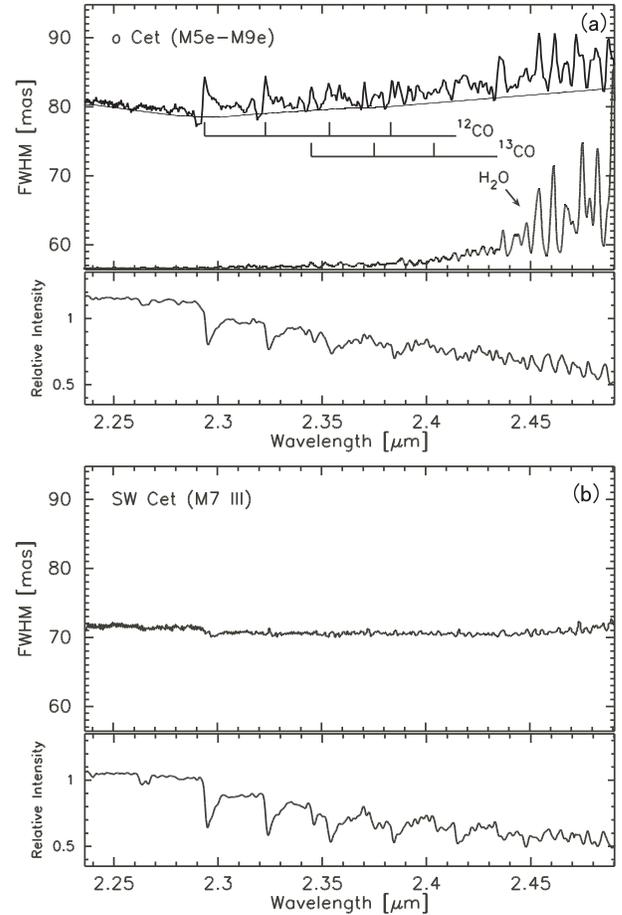}
 \end{center}
\caption{(a) The FWHM of the spatial profile as a function of wavelength for $o$~Cet is presented in the upper panel and the intensity spectrum is shown in the lower panel. The FWHM is larger by up to 10\% at the molecular lines relative to the continuum wavelengths. The enhancement in the FWHM closely follows the molecular absorption features of H$_2$O and CO. For comparison, the pure H$_2$O opacity at 2000~K computed with the HITRAN/HITEMP database (arbitrary units) is given at the bottom of the upper panel \citep{rot98}. There is good correlation between the opacity plot and the measured stellar size at strong water vapor absorption bands longer than 2.42~\micron. The spectrum of the opacity plot is smoothed with the spectral resolution R = 1000 of our observation. 
We have fitted the lower envelope of the observed FWHM using a polynomial function $\theta_{\rm obs\_fit}(\lambda)$. We regard the lower envelope as the apparent stellar size of the photosphere without molecular lines convolved with the system PSF. (b) Same as (a), but for the comparison star SW~Cet. Note that no FWHM enhancement is visible for this star. The uncertainty of the measurement is estimated from the variation in (b) to be approximately +/- 1 mas. \label{f2}}
\end{figure}

Figure~\ref{f1} shows the spectra and spatial profiles of $o$~Cet, the target star, and SW~Cet, the reference unresolved star. The spatial extent of $o$~Cet is slightly greater at the molecular absorption lines than at the adjacent continuum wavelength. In contrast, no variation is seen in the spatial profile of SW~Cet, measured in the same way as that of $o$~Cet. In Figure~\ref{f2}, we see that the increase in the measured FWHM of $o$~Cet is well correlated with the H$_2$O opacity feature in the 2.43-2.48~\micron~region and the CO bandhead in the 2.29--2.41~\micron~region. This is clear evidence for extension of the water vapor and CO atmospheres outside the photosphere of $o$~Cet. 
Note that the FWHM of the reference star SW~Cet, an unresolved point source, is uniform and is not correlated with the absorption line features. 
It shows that a difference in diameter much smaller than the point spread function can be measured with this differential spectral imaging technique. 

The stellar radius is smaller than the continuum at the blue end of the CO bands at 2.29 and 2.32~\micron. These do not seem to be real features, but we do not have a clear explanation for their origin. One possible explanation is that they could be artifacts caused by the sharp drop in flux at these wavelengths. Note, however, that this effect is very small in the SW~Cet data with similar CO absorption features (Figure~\ref{f2}(b)).

The size of the extended gas is estimated from the current data as follows. The measured apparent stellar size is the convolution of the real size with the diffraction-limited PSF at 2.2~\micron~for an 8.2-m telescope. Deconvolving the profile to evaluate the real stellar size directly from AO observations is not straightforward, because the PSF of the AO observations is time varying. Therefore the deconvolved stellar profile from a comparison star is not usually correct. To overcome the time variation problem, we compare the stellar size at the wavelength of the molecular lines to that at the continuum wavelength in the same observation of Mira. That is, we use the stellar profile at the continuum wavelength as the simultaneous PSF reference. By adopting the photosphere size derived from other interferometric observations in the continuum, we can derive the stellar size measured at the molecular lines.

Although the center-to-limb variation (CLV) of the intensity across the stellar disk has a complex profile, depending on the stellar model used \citep{sch01,jac02}, we assumed a Gaussian function for simplicity. The Gaussian CLV is sufficient to estimate the FWHM of the molecular layer because we do not have any higher spatial resolution information. The apparent spatial width is then a square sum of the PSF size and the angular size of the star,

\[
\theta^2_{\rm obs}(\lambda) = \theta^2_{\rm \ast}(\lambda)
+ \theta^2_{\rm PSF}(\lambda).
\]

We need to know the size of the PSF as a function of the wavelength, $\theta_{\rm PSF}(\lambda)$, to calculate the stellar diameter. However, the PSF measured with the comparison star in this observation is not useful because it can differ from that obtained during observation of the object due to changes in the seeing conditions. Instead, we estimate the system PSF using the following procedure. We assume a stellar photosphere without any molecular layer to be the size given by $\theta_{{\rm \ast}0}$. The photosphere defines the stellar diameter at all wavelengths where there are no molecular lines. The derived photospheric stellar size, $\theta_{\rm obs\_fit}(\lambda)$, should have the following relationship with the system PSF.

\[
\theta^2_{\rm obs\_fit}(\lambda) = \theta^2_{{\rm \ast}0}(\lambda)
+ \theta^2_{\rm PSF}(\lambda).
\]

From Figure~\ref{f2}, we obtained $\theta_{\rm obs\_fit}(\lambda)$ by fitting the lower envelope of the observed FWHM with a polynomial function. The lower envelope is our approximation to the continuum level of the star spectrum. The PSF size is then calculated with $\theta^2_{\rm PSF}(\lambda) = \theta^2_{\rm obs\_fit}(\lambda) - \theta^2_{{\rm \ast}0}(\lambda)$. We assume that the intrinsic diameter of the stellar photosphere $\theta_{{\rm \ast}0}(\lambda)$ is constant in the relevant wavelength region. The diameter $\theta_{{\rm \ast}0}$ of the photosphere at 2.22~\micron, 20.8~+/-0.28~mas, was provided by fitting the visibility data of interferometric observation by \citet{per04}. Their observation (phase $\phi$=0.20) was made one month prior to ours ($\phi$=0.30). Since \citet{per04} found only a 3\% increase in the photospheric diameter from $\phi$=0.01 to $\phi$=0.2, we assume the same photospheric diameter.
Since we normalize to the photospheric diameter, the absolute value is not critical. Then, the stellar diameters including molecular lines are

\[
\theta_{\rm \ast}(\lambda) = \sqrt{ \theta^2_{\rm obs}(\lambda)
- \left[ \theta^2_{\rm obs\_fit}(\lambda)
- \theta^2_{{\rm \ast}0} \right]}.
\]

\begin{figure}[p]
\begin{center}
\FigureFile(80mm,90mm){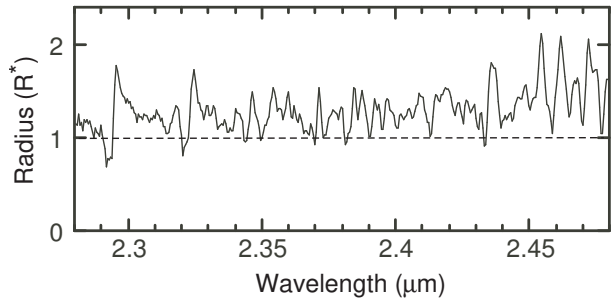}
\end{center}
\caption{Observed radius relative to the photospheric radius of $o$~Cet. The radius of the star extends up to twice the size of the photosphere at the H$_2$O and CO lines. The estimation is a lower limit of the extension. The error of the radius from the overlapping molecular lines is estimated to be 10\%. The effect of the low spectral resolution observation will be confirmed by observation with higher spectral resolution. \label{f3}}
\end{figure}

Figure~\ref{f3} presents the stellar size so derived as a function of wavelength, and shows that for $o$~Cet, the molecular layers of water vapor and CO extend to twice the photospheric radii. 
We have examined the error in deriving the enhancement of the molecular layer radius from the photosphere. The error sources are the combination of random and systematic error. The random error is estimated to be +/- 1\%, which comes from the noise level of the radius in the wavelengths of no molecular features. 
For the systematic error, we have identified that the moderate spectral resolution is a major source. 
 The size of $\theta_{\rm obs\_fit}(\lambda)$ might be an overestimate because we do not resolve the individual molecular lines by our spectral resolution of R = 1000. 
The bottom level of the molecular opacity curve smoothed with this spectral resolution is not zero. This means that the apparent continuum size we have derived by lower envelope fitting gives a larger value than the actual size. This would lead to an underestimate of the extent of the molecular atmosphere. We have estimated this effect for the lines between 2.45 and 2.48~\micron. When we look at the model opacity curve at this wavelength, the real apparent continuum size will be about 80~mas, while we set the lower envelope at this wavelength to 82.5~mas in Figure~\ref{f2}. This changes the estimated stellar size from 2.12 to 2.33 R*. The uncertainty of the radius due to this effect is about 10\%. 

The lower spectral resolution observation also provides another underestimate of the extent of the molecular layer, because the peaks of the opacity are also smoothed out and do not reach the peak positions of the individual lines. 

The error of continuum diameter of Mira taken from \citet{per04} is a source of the error in calculating the molecular layer diameter.  From the above formula, the error of molecular diameter is estimated to be the same order of error of the continuum diameter, which is, approximately 1\%.

We compared our result to previous studies.
The H$_2$O atmosphere of $o$~Cet measured using the ISO spectroscopic observations \citep{yam99} extends to 2.3~$R_{\rm \ast}$ ($\phi$=0.99). \citet{wei04} modeled the gas layer from the spectroscopic and interferometric data and derived a diameter of 2.4~$R_{\rm \ast}$ ($\phi$=0.81). \citet{per04} got 2.0~$R_{\rm \ast}$ at 2.39~\micron~ from their optical interferometer observations with IOTA. These observations gave numbers similar to our measurements.
\citet{ire04} reported a diameter of 1.5~$R_{\rm \ast}$ at 2.46~\micron~using Keck aperture masking observations, a value that is significantly smaller than our result.
However, their spectral resolution was much lower than ours.

Observations at higher spectral resolution should reduce these uncertainties and provide improved measurement for this type of study. The signal-to-noise ratio of the Subaru Telescope is sufficiently high to make observations at a resolution ten times greater with the IRCS using this method.

\section{Summary}
We observed $o$~Cet with adaptive optics in the {\it K}-band with diffraction-limited spatial resolution. Using the differential spectral imaging technique, we confirmed that the stellar size measured at the molecular absorption lines is significantly larger than that measured at the adjacent continuum wavelengths.

The effective diameters of the H$_2$O and CO atmospheres of $o$~Cet are about twice that of the photosphere observed in the continuum. 
This is the first direct spectroscopic imaging of the extended molecular atmosphere of a late-type star with sufficient spectral resolution to resolve those lines.
The size of the atmosphere measured is consistent with infrared interferometric observations and the spectroscopic study of the star by ISO.

\bigskip
We thank all the staff and crew of the Subaru Telescope and NAOJ for
their valuable assistance in obtaining these data and continuous support
for the Subaru AO and IRCS construction. Miwa Goto is supported by a Japan Society for the Promotion of Science fellowship.

\end{document}